\documentclass[aps,prd,twocolumn,fleqn,superscriptaddress,preprintnumbers]{revtex4}

\usepackage{graphicx,color}
\usepackage{amsmath,amssymb,amsfonts}

\newcommand{\be}{\begin{equation}}
\newcommand{\ee}{\end{equation}}
\newcommand {\nn} {\nonumber}

\newcommand{\bea}{\begin{eqnarray}}
\newcommand{\eea}{\end{eqnarray}}

\newcommand{\Ein}{E_\mathrm{in}}
\newcommand{\Eout}{E_\mathrm{out}}

\newcommand{\ba}{\begin{eqnarray}}
\newcommand{\ea}{\end{eqnarray}}
\renewcommand{\(}{\left(}
\renewcommand{\)}{\right)}

\newcommand{\w}{\hat{\omega}}

\def\be{\begin{equation}}
\def\ee{\end{equation}}
\def\beq{\begin{eqnarray}}
\def\eeq{\end{eqnarray}}

\begin{document}

\input amssym.def
\input amssym.tex

\title{Production of Prompt Photons: \\
 Holographic Duality and Thermalization}

\preprint{BI-TP 2012/24, TUW-12-13}

\author{Rudolf Baier}
\affiliation{Faculty of Physics, University of Bielefeld,
D-33615 Bielefeld, Germany}
\author{Stefan A. Stricker}
\affiliation{Institut f\"{u}r Theoretische Physik,
Technische Universit\"{a}t Wien,\\
Wiedner Hauptstr.~8-10, A-1040 Vienna, Austria}
\author{Olli Taanila}
\affiliation{Faculty of Physics, University of Bielefeld,
D-33615 Bielefeld, Germany}
\author{Aleksi Vuorinen}
\affiliation{Faculty of Physics, University of Bielefeld,
D-33615 Bielefeld, Germany}

\begin{abstract}
We study the production of prompt photons in strongly coupled out-of-equilibrium Super Yang-Mills plasma using the AdS/CFT correspondence. Our goal is to determine the photon emission spectrum at different stages of a thermalization process, which is modeled via the gravitational collapse of a thin spherical shell in AdS$_5$ space. Particular emphasis is placed on the limit of large frequencies, which we are able to treat analytically.
\end{abstract}

\maketitle

{\em Introduction and setup.}
In heavy ion experiments highly energetic electromagnetic probes, such as photons and dileptons, provide important information on the evolution and eventual thermalization of the produced plasma (for reviews, see e.g.~\cite{Arleo,Gale}). This is due in particular to their weak coupling to the plasma constituents, which implies that once produced, they are free to propagate through the plasma almost unaltered. The spectrum of direct photons measured in central {\it Au-Au} collisions indeed shows an enhancement above the scaled {\it p-p} spectrum for transverse momenta less than 2 GeV  \cite{Adare:2008ab,Reygers:2009tm,Sakaguchi:2010hx}. This appears to be consistent with the production of thermal photons in a strongly coupled quark-gluon plasma, as indicated by both RHIC \cite{Tannenbaum} and LHC experiments \cite{Muller}.

On the theory side, the dynamics of strongly interacting field theories out of thermal equilibrium is a notoriously difficult problem, as perturbative results are only applicable in the limit of asymptotically high energies, while lattice methods are in general constrained to equilibrium quantities \cite{Arnold,Meyer}. In this context, the gauge/gravity duality has proved itself highly useful, as it allows one to reduce the strongly coupled dynamics of certain field theories to classical gravity problems in curved spacetime \cite{Maldacena,Gubser,Witten}. Thermalization on the field theory side is argued to be dual to the formation of a black hole in an (asymptotically) Anti de Sitter (AdS) spacetime, which one can model e.g.~via the gravitational collapse of a thin shell of matter in this geometry \cite{Danielsson1,Danielsson2,LinShuryakIII,Balasubramanian,Rudolf,Wu:2012ri}.

In the present paper we concentrate on the production of prompt photons in a thermalizing plasma, aiming to extend our previous work on dileptons \cite{Rudolf}. In particular, we are interested in the (isotropic) differential production rate of on-shell photons with energy $k^0=k$ \cite{Huot,Myers},
\be\label{rate}
k^0 \frac{d\Gamma_{\gamma}}{d^3k} = \frac{1}{4\pi k}\frac{d\Gamma_{\gamma}}{dk_0} =
\frac{\alpha}{4 \pi^2} \eta^{\mu\nu} \Pi_{\mu \nu}^{<} (k^0 = k) ~,
\ee
where $\alpha$ is the fine structure constant and $\Pi_{\mu \nu}^{<}$ the electromagnetic current Wightman function. In thermal equilibrium, the fluctuation dissipation theorem allows one to further relate $\Pi_{\mu \nu}^{<}$ to the (transverse) photon spectral function $\chi_{\mu\nu}$,
\be
\eta^{\mu\nu} \Pi_{\mu \nu}^{<} (k^0 = k) \, = \,n_B(k^0) \chi_{\mu}^{\mu}(k_0)\,\equiv\, n_B(k^0) \chi(k_0)  ~.\label{wight}
\ee
While this is not generically true out of equilibrium, it was shown to hold in the quasistatic limit of the falling shell system in appendix B of \cite{Rudolf}.

The gravity setup we work in is thoroughly explained in \cite{Rudolf}, and only briefly summarized here. We place an infinitesimally thin shell of unspecified matter at some radius $r_s$ in AdS$_5$ space, and let it fall gravitationally in the radial direction. The limit where $r_s$ approaches the Schwarzschild radius $r_h$ of the shell is conjectured to correspond to thermalization in strongly coupled, large-$N_c$ $\mathcal{N}=4$ Super Yang-Mills (SYM) plasma, while $r_h$ (and thus the mass of the shell) is related to the final equilibrium temperature of the field theory.

Next, we introduce in the SYM theory a U(1) gauge field coupled to the conserved current corresponding to a U(1) subgroup of the SU(4) R-symmetry. According to the standard AdS/CFT prescription \cite{SonStarinets}, the corresponding retarded correlator --- and thus the `photon' spectral function --- is calculable by studying linearized fluctuations of an (unrelated) U(1) gauge field in the bulk, where particular care must be taken of the boundary conditions of the field at the shell \cite{Rudolf}. We perform the calculation in the so-called quasistatic approximation \cite{Danielsson1}, in which the shell is assumed to move slowly compared to the other time scales of interest. As discussed in section 4 of \cite{Rudolf}, this condition is satisfied in the later stages of thermalization as well as for highly energetic photons, with frequency larger than the inverse timescale associated with the motion of the shell.

Finally, during the past few months, a number of related works on the subject of holographic thermalization have appeared on the arxiv; for details, we refer the interested reader to  refs.~\cite{Chesler,Galante,Erdmenger,Caceres,Mukhopadhyay} and references therein.

{\em The metric and the equation of motion.}
The metric of the falling shell setup consists of a black hole solution outside the shell, $r > r_s$, and of pure AdS space inside it, $r<r_s$. The parameter $r_s$ is assumed to be larger than the Schwarzschild radius of the shell, $r_s>r_h$, so that initially there is no black hole and the dual system is out of equilibrium. Assuming $r_h$ to be much larger than the curvature radius of $AdS$ space, $r_h\gg L$, the line element of the spacetime can be written in the form
 \begin{equation}\label{fullAdS}
ds^2 = \frac{r^2}{L^2}\left( -f(r) \, dt^2 + d\mathbf{x}^2\right)
+ \frac{L^2}{r^2}\frac{dr^2}{f(r)}\;,
\end{equation}
where we have defined
\be
f(r) \,=\, \left\{ \begin{array}{lr}
1- \frac{r_h^4}{r^4} \, ,& \mathrm{for}\;r >r_s\\
1\, ,& \mathrm{for}\; r < r_s
\end{array}\right. \;.
\ee
The Hawking temperature of the eventual black hole is given by $T=\frac{r_h}{\pi L^2}$; this is also the temperature in the field theory that we will be referring to throughout the evolution of the system (although it is in principle well-defined only at the end). For notational convenience, we will in the following set $L=1$ and introduce a new radial variable $u \equiv \frac{r_h^2}{r^2}$. In this new coordinate, the boundary is located at $u=0$ and the horizon at $u=1$, while the only relevant parameter of the gravity setup, the shell location, takes the form $u_s=\frac{r^2_h}{r^2_s}$.

To study photon production, we need to investigate a transverse electric field $E^{\perp}(t,u,\mathbf{x})$ \cite{Kovtun,Myers}. Denoting $k_0=\omega$ and using the fact that for an on-shell photon $k=k_0$, we obtain for its equation of motion in momentum space \cite{Huot}
\begin{equation}\label{eomEz}
\partial^2_u E^{\perp} + \frac{\partial_u f}{f}\partial_u E^{\perp}
 +\frac{{\hat{\omega}}^2 u}{f^2}E^\perp = 0\;
 ,\quad\partial_u \equiv \frac{\partial}{\partial u} \; ,
\end{equation}
where we have introduced the dimensionless variable ${\hat{\omega}} \equiv \frac{k^0}{2r_h} = \frac{k^0}{2\pi T}$. This equation has two linearly independent solutions, which are given by hypergeometric functions \cite{Abramowitz} (see also \cite{Myers})
\begin{eqnarray}
\label{Esolin}
\hspace{-0.7cm} \Ein^{\perp}(\hat{\omega},u) &=&
 (1-u)^{-\frac{i{\hat{\omega}}}{2}}(1+u)^{-\frac{{\hat{\omega}}}{2}} \times \\ \nn
&&\hspace{-1.2cm}{}_2F_1\left(1-\frac{1+i}{2}{\hat{\omega}},-\frac{1+i}{2}{\hat{\omega}};
1-i{\hat{\omega}};\frac{(1-u)}{2}\right)\, ,\\
\hspace{-0.5cm}\label{Esolout}\Eout^{\perp}(\hat{\omega},u) &=& (1-u)^{\frac{i{\hat{\omega}}}{2}}
(1+u)^{-\frac{{\hat{\omega}}}{2}}\times \\
\nn
&&\hspace{-1.2cm} {}_2F_1\left(1-\frac{1-i}{2}{\hat{\omega}},-\frac{1-i}{2}{\hat{\omega}};
1+i{\hat{\omega}};\frac{(1-u)}{2}\right)\, ,
\end{eqnarray}
where $\Ein^{\perp}$ satisfies an infalling and $\Eout^{\perp}$ an outgoing boundary condition at the horizon.

In a black hole background, the physical solution is the one obeying infalling boundary conditions, as classically nothing can escape from a black hole \cite{SonStarinets,Kovtun}. In the case of a falling shell, there is, however, no horizon, and hence we must write our full solution as a linear combination of the infalling and outgoing components,
\begin{equation}
E^{\perp}_\mathrm{outside} (\hat{\omega},u) = c_+ \Ein^{\perp}(\hat{\omega},u)
 +c_-\Eout^{\perp}(\hat{\omega},u) \;, \label{Eoutdefn}
\end{equation}
and match this to the solution inside the shell. As discussed in \cite{Rudolf}, the discontinuity of the time coordinate across the shell implies that the frequency outside the shell is related to the inside one through a rescaling by $\sqrt{f_m}\equiv \sqrt{1 - u_s^2}$. Taking this into account, the equation of motion inside the shell is given by
\beq\label{ins}
\partial^2_u E^{\perp}_{inside} + \frac{\hat{\omega}^2  (1 - f_m)}{f_m u}
E^{\perp}_{inside} = 0 \; ,
\eeq
the solution of which reads in terms of Bessel functions
\beq
E_\mathrm{inside} (\hat{\omega},u) &=& 
\sqrt{u}\left[ J_1\left(2\tilde{\omega}\sqrt{u}\right) + 
i Y_1 \left(2\tilde{\omega}\sqrt{u}\right)\right] \nn \\
 &\stackrel{u\to 0}{=}&
\frac{1}{\sqrt{\tilde{\omega}}u^{1/4}}~
e^{i \tilde{\omega}\sqrt{u}} \; ,
\eeq
where we have denoted
\be
\tilde{\omega} \equiv\hat{\omega}\,\sqrt{\frac{1-f_m}{f_m}}\; .
 \ee
The matching of the two solutions then leads to the result (for details, see \cite{Rudolf})
\be\label{Cmp}
\frac{c_-}{c_+} = -\frac{\Ein^{\perp}\partial_u E^{\perp}_\mathrm{inside}-\sqrt{f_m}
   E^{\perp}_\mathrm{inside}
\partial_u \Ein^{\perp}}{\Eout^{\perp}\partial_u
E^{\perp}_\mathrm{inside}
-{\sqrt{f_m}} E^{\perp}_\mathrm{inside}\partial_u \Eout^{\perp}}\bigg|_{u=u_s} \, .
\ee
In the limit $u_s\to 1$, this ratio vanishes as
\beq\label{limit}
\frac{c_-}{c_+} \rightarrow \frac{i(1-u_s)^{1/2-i\w}}{4\sqrt{2}\w}\rightarrow 0
\eeq
which serves as a consistency check of our calculation (note that we are only interested in real values of $\hat{\omega}$ here).

\begin{figure}
\centering
\includegraphics[width=8.2cm]{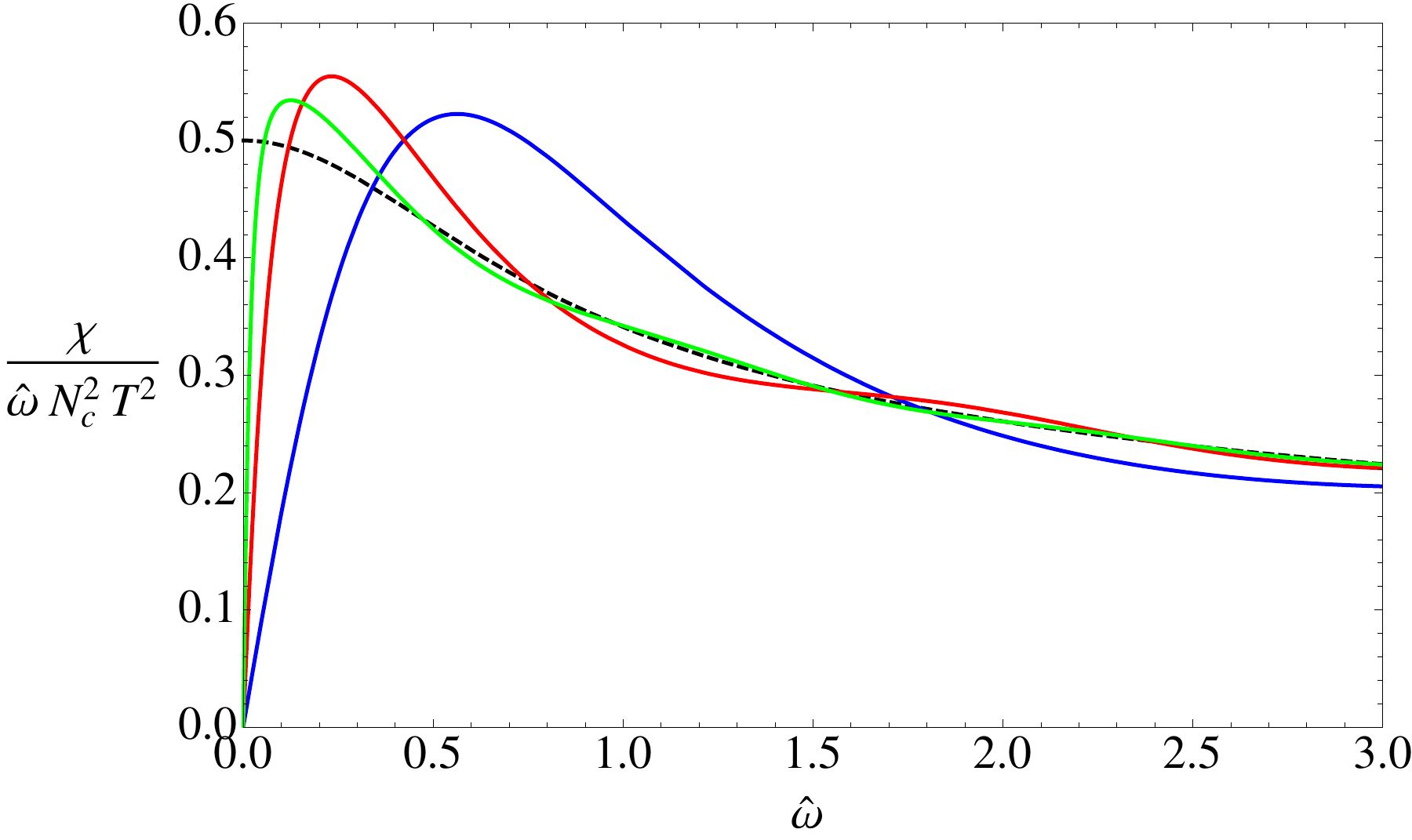}
\caption {The trace of the spectral function $\chi(\hat{\omega},u)$, normalized by $\hat{\omega} N_c^2 T^2$. In order of increasing amplitude of the fluctuations, the curves correspond to $r_s/r_h=1.001$ (green), 1.01 (red) and 1.1 (blue). The dashed black line stands for the thermal spectral density.}
\label{rho}
\end{figure}

{\em Photon production.}
From eqs.~(\ref{rate})--(\ref{wight}), we see that in order to study the production of prompt photons, it suffices to evaluate the trace of the transverse photon spectral function. As the latter is given by the imaginary part of the retarded R symmetry correlator $\Pi^\perp$, we can follow the standard AdS/CFT prescription of \cite{SonStarinets} and write
\ba\label{spectral}
\chi(\hat{\omega})&=&-4\,  \mathrm{Im}\, \Pi^{\perp} (\hat{\omega}) \nn\\
&=& \frac{N_c^2 T^2}{2}
\mathrm{Im} \frac{\partial_u E^{\perp}_\mathrm{outside} (\hat{\omega}, u)}{E^{\perp}_\mathrm{outside}(\hat{\omega},u)} \Big|_{u =0}~,
\ea
where we have used the shorthand $\phi(\omega=2\pi T \hat{\omega})\equiv \phi(\hat{\omega})$ for the functions $\chi$ and $\Pi^\perp$.  This quantity is furthermore conveniently expressed in terms of the Wronskian of
\ba
F^{\perp}(\hat{\omega},u) &\equiv& \frac{E^{\perp}_\mathrm{outside}(\hat{\omega}, u)}{E^{\perp}_\mathrm{outside}(\hat{\omega}, u=0)}~,
\ea
producing \cite{Huot}
\begin{eqnarray}\label{Wrons}
\hspace{-0.5cm}\mathrm{Im}\,\Pi^{\perp} (\hat{\omega}) & \equiv& - \frac{N_c^2 T^2}{8} W(\hat{\omega}, u)\, , \\
W(\hat{\omega}, u)& \equiv&
 - \frac{N_c^2 T^2}{8}  \mathrm{Im}\,\Big[f(u) (F^{\perp}(\hat{\omega},u))^{*}
 \partial_u F^{\perp}(\hat{\omega},u)\Big] ~. \nonumber
\end{eqnarray}
The advantage of this formulation is that it can be easily shown that $W(\hat{\omega}, u)$ does not depend on the variable $u$, i.e.~$W(\hat{\omega}, u) = W(\hat{\omega})$. Thus, we can determine its value in the numerically more convenient limit $u\rightarrow 1$ instead of the boundary, $u=0$.

A lengthy but straightforward calculation employing the formalism explained above leads us to the result
\ba
\label{chi}
\chi(\hat{\omega},u_s) &=& \frac{N_c^2 T^2 \hat{\omega}}{8}\frac{1 - \vert \frac{c_-}{c_+}\vert^2}{D(\hat{\omega})} \; ,
\ea
where we have reinstated the explicit $u_s$ dependence of the spectral function (originating from $c_\pm$) as well as denoted
\ba\label{Fom}
\hspace{-0.7cm}D(\hat{\omega}) &\equiv&  \vert {}F \lvert^2  \(1 + \vert \frac{c_-}{c_+} \rvert^2 \)\!+2\,\mathrm{Re}\left[F^2\! \(2^{i\hat{\omega}} \,\frac{c_-}{c_+}\)^{*} \right] \!, \\
\hspace{-0.7cm} F &\equiv& {}_2F_1 \left( 1-\frac{1+i}{2}{\hat{\omega}},1 + \frac{1-i}{2}{\hat{\omega}}; 1-i{\hat{\omega}};-1 \right)  . \nonumber
\ea
From eq.~(\ref{limit}), we know that the ratio $c_-/c_+$ approaches $0$ when $u_s\to 1$. In this limit, the spectral function thus reduces to its known equilibrium limit,
\ba
\label{thermal}
\chi_{\text{thermal}}(\hat{\omega}) &=&
\frac{N_c^2 T^2 \hat{\omega}}{8} \frac{1}{\vert  F \vert^2} ~,
\ea
as required by consistency (cf.~eq.~(3.18) of \cite{Huot}).

\begin{figure}
\centering
\includegraphics[width=7.8cm]{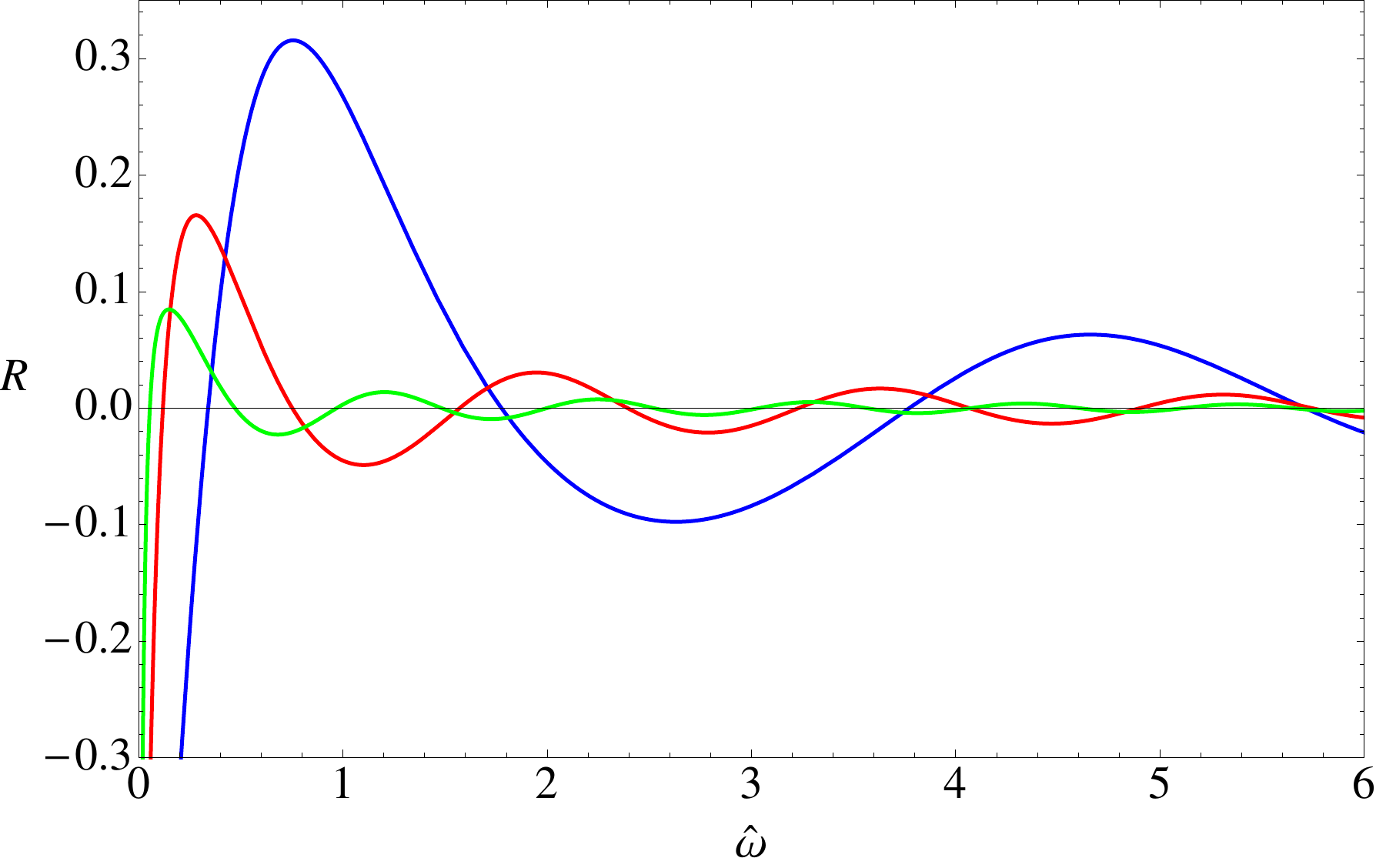}
\caption {The relative deviation of the photon spectral function from its thermal limit, $R(\hat{\omega},u_s)$, for $r_s/r_h=1.001,\,1.01,\;1.1$. The color coding is as in fig.~\ref{rho}.}
\label{Rphoton}
\end{figure}

The novel aspect of our result is that eq.~(\ref{chi}) gives the photon spectral density both in and out of thermal equilibrium. As thermalization is parameterized by the shell location, we use several different values of $u_s$ when plotting this function in fig.~\ref{rho}. In fig.~\ref{Rphoton}, we also show the relative deviation of the spectral density from its thermal limit,
\ba
\label{R}
R(\hat{\omega},u_s) &\equiv& \frac{\chi(\hat{\omega},u_s)-\chi_\text{thermal}(\hat{\omega})}{\chi_\text{thermal}(\hat{\omega})} \, ,
\ea
which exhibits an oscillatory pattern much like the one already observed in the dilepton spectral function in \cite{Rudolf} (for a discussion of the physical origin of the oscillations, see the same reference). As the shell approaches the horizon, $r_s\to r_h$, the amplitude of these oscillations is seen to decrease, as it of course should. The amplitude is also seen to be dampened at large values of $\hat{\omega}$, which we interpret as a sign of the usual top/down type pattern of holographic thermalization. In the opposite limit of small $\hat{\omega}$, the out-of-equilibrium $\chi/\hat{\omega}$ is on the other hand seen to vanish; this is, however, merely a sign of departing the range of validity of the quasistatic approximation.

\begin{figure}
\centering
\includegraphics[width=8.6cm]{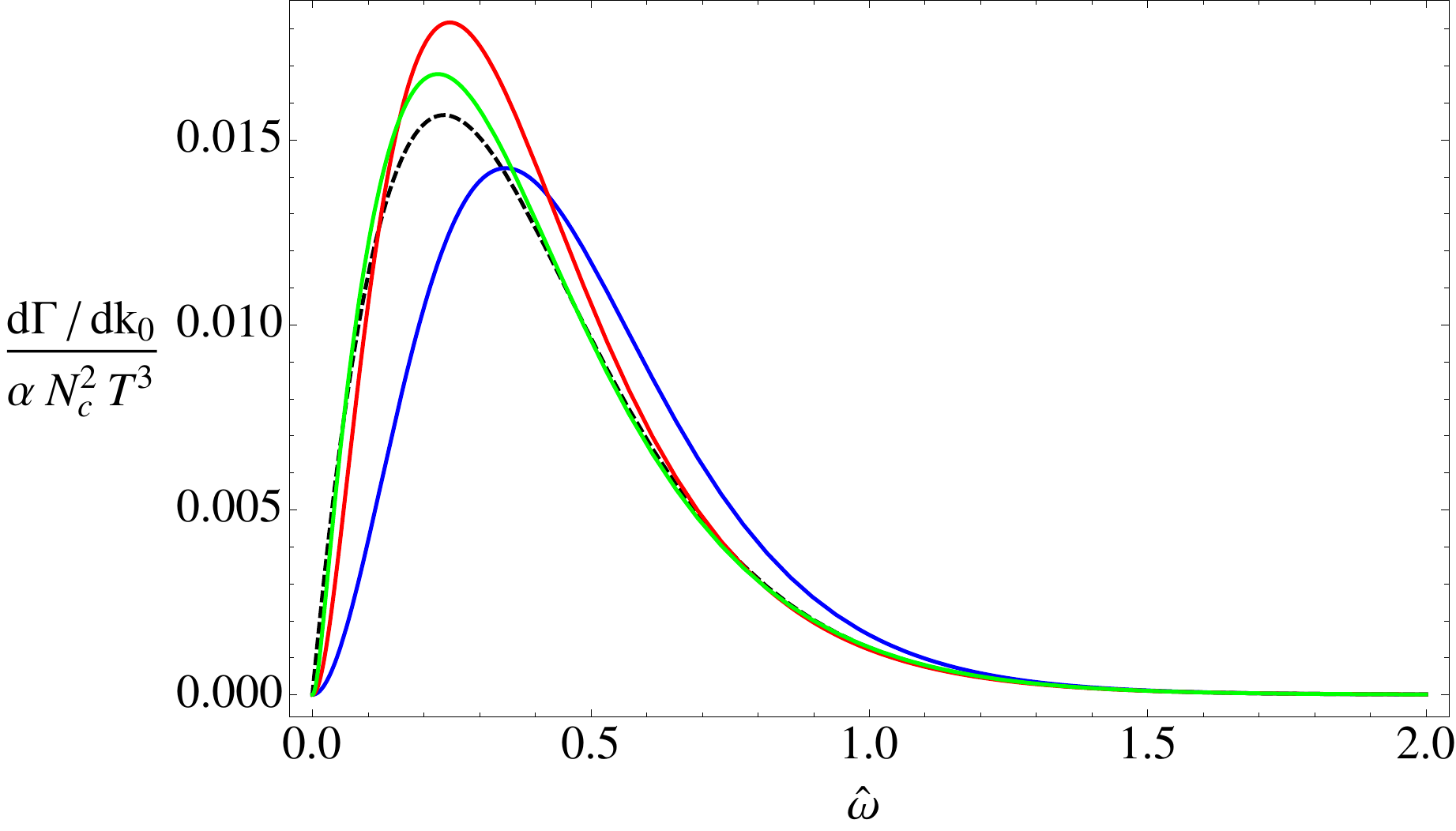}
\caption {The photon emission spectrum $d\Gamma_\gamma/k_0$, normalized by $\alpha N_c^2 T^3$, for $r_s/r_h=1.001$, 1.01 and 1.1. The color coding is as in fig.~\ref{rho}.
}
\label{photonrate}
\end{figure}

In fig.~\ref{photonrate} we finally show the photon emission rate per unit volume as a function of $\hat{\omega}$,
\begin{equation}
\frac{d\Gamma_\gamma}{dk_0}=\frac{\alpha k}{\pi}n_B(k_0) \chi(k_0)\Big|_{k_0=k=2\pi T\hat{\omega}}\, .
\end{equation}
We observe a slight enhancement of the photon production rate, when the system is out of thermal equilibrium. Nearly all traces of the oscillatory behavior have, however, been suppressed by the Bose-Einstein distribution function, and in particular the overall shape of the spectrum stays largely unaltered when $r_s>r_h$. It is interesting to contrast this behavior with that of the leading $1/\lambda^{3/2}$ corrections to the equilibrium photon emission rate studied in \cite{Hassanain}. While the equilibrium emission spectrum of \cite{Huot} was seen to get a moderate enhancement also in this case, it was found that the peak of the function moved towards the IR with decreasing $\lambda$, while in our case it shifts towards the UV.

{\em Large $\hat{\omega}$ limit.}
In analogy with the dilepton case of ref.~\cite{Rudolf}, it would clearly be of some value to get an analytic handle on the behavior of the photon spectral density in the limit of large $\hat{\omega}$, where the quasistatic approximation is known to work. To this end, we follow the discussion of appendix A of \cite{Huot} and first note that for large $\hat{\omega}$, the general solution to the equation of motion (\ref{eomEz}) can be expressed in terms of Airy functions,
\ba
\hspace{-0.4cm} E^{\perp}_\mathrm{outside} (\hat{\omega},u)&=&\frac{h^{-1/4}(-u)}{\sqrt{-f(u)}}\bigg\{C_1(\hat{\omega}){\rm Ai}(\hat{\omega}^{2/3}\zeta(-u))\nn \\ &+&C_2(\hat{\omega}){\rm Bi}(\hat{\omega}^{2/3}\zeta(-u))\bigg\}, \label{Eperp}
\ea
with $C_i(\hat{\omega})$ denoting unknown coefficient functions and
\ba
\label{zetay} \zeta(y)&\equiv& e^{-i\pi/3} \bigg\{\frac{3\, g(-y)}{4}\bigg\}^{2/3}\, , \\
g(y)&\equiv &i\ln\frac{\sqrt{-y}+1}{\sqrt{-y}-1}-2i\,{\rm arctan}(\sqrt{-y})-\pi\, , \\
h(y)&\equiv&\frac{y}{(1-y^2)^2\zeta(y)}\, .
\ea
One should note here that the quantity inside the curly brackets in eq.~(\ref{zetay}) is real and positive in the range $0\leq y \leq 1$.

It is a straightforward exercise to show that close to the horizon, $u\approx 1$, eq.~(\ref{Eperp}) reduces to
\ba
&&\hspace{-0.8cm}E^{\perp}_\mathrm{outside} (\hat{\omega},u)\to \frac{e^{-i\pi/4}}{2\sqrt{\pi}\hat{\omega}^{1/6}} \bigg\{C_1(\hat{\omega})2^{i\hat{\omega}}e^{-i\pi \hat{\omega}/4}(1-u)^{-i\hat{\omega}/2} \nonumber \\
&&\hspace{1.9cm}+C_2(\hat{\omega})\Big(2^{1-i\hat{\omega}}e^{i\pi \hat{\omega}/4}(1-u)^{i\hat{\omega}/2}\nn\\
&&\hspace{1.9cm}-i\,2^{i\hat{\omega}}e^{-i\pi \hat{\omega}/4}(1-u)^{-i\hat{\omega}/2}\Big)\bigg\} \nn \\
&=& \frac{2^{\hat{\omega}/2}e^{-i\pi/4}}{2\sqrt{\pi}\hat{\omega}^{1/6}}
\bigg\{2^{i\hat{\omega}}e^{-i\pi\hat{\omega}/4}(C_1(\hat{\omega})-iC_2(\hat{\omega}))\Ein(\hat{\omega},u)\nonumber \\
&&\hspace{1.9cm}+2^{1-i\hat{\omega}} e^{i\pi \hat{\omega}/4}C_2(\hat{\omega})\Eout(\hat{\omega},u)\bigg\}\, ,
\ea
where in the latter equality we have used the $u\to 1$ limit of eqs.~(\ref{Esolin})--(\ref{Esolout}). Denoting finally
\ba
\alpha&\equiv& \frac{2^{\hat{\omega}/2}e^{-i\pi/4}}{2\sqrt{\pi}\hat{\omega}^{1/6}}\, ,\quad\quad
\beta \,\equiv \, 2^{i\hat{\omega}}e^{-i\pi\hat{\omega}/4}\, ,
\ea
we then obtain upon comparison with eq.~(\ref{Eoutdefn}) the coefficients $C_i(\hat{\omega})$,
\ba
C_1(\hat{\omega})&=&\alpha^{-1}\beta^{-1}c_+ +\frac{i}{2}\alpha^{-1}\beta c_-\, , \\
C_2(\hat{\omega})&=&\frac{1}{2}\alpha^{-1}\beta c_-\, .
\ea

A very useful byproduct of the above calculation is that we can now write the properly normalized infalling and outgoing solutions to the equation of motion (\ref{eomEz}) in the large-$\hat{\omega}$ limit as
\ba
\Ein^\perp(\hat{\omega},u)&=&\frac{h^{-1/4}(-u)}{\alpha\beta\sqrt{-f(u)}}\,{\rm Ai}(\hat{\omega}^{2/3}\zeta(-u))\, ,  \\
\Eout^\perp(\hat{\omega},u)&=&\frac{\beta h^{-1/4}(-u)}{2\alpha\sqrt{-f(u)}}\Big(i{\rm Ai}(\hat{\omega}^{2/3}\zeta(-u))\\
&&+{\rm Bi}(\hat{\omega}^{2/3}\zeta(-u))\Big)\nn\, ,
\ea
from which we further obtain
\ba\label{Elarge}
\hspace{-0.6cm}E^\perp_{\substack{\mathrm{in}\\\mathrm{out}}} (\hat{\omega},u)\stackrel{\hat{\omega}\to \infty}{\to} 2^{-\hat{\omega}(1\pm 2i)/2}e^{\pm i\hat{\omega}(\pi+2g(u))/4}u^{-1/4}\, .
\ea
This is helpful in determining the large-$\hat{\omega}$ limit of the retarded Green's function of interest,
\ba
\hspace{-0.9cm}\Pi^\perp_\text{asym}(\hat{\omega},u_s)&=&-\frac{N_c^2T^2}{8}\frac{3^{1/3}\Gamma(2/3)}{\Gamma(1/3)}\label{PiTas}\\
&&\times
\frac{1-2^{2i\hat{\omega}}e^{-i\pi/2(\hat{\omega}+1/3)}\frac{c_-}{c_+}}{1+2^{2i\hat{\omega}}e^{-i\pi/2(\hat{\omega}-1/3)}\frac{c_-}{c_+}} (-\hat{\omega})^{2/3}\, , \nn 
\ea
which clearly reduces to the correct equilibrium result of \cite{Huot} when $c_-/c_+\to 0$.

\begin{figure}
\centering
\includegraphics[width=8.0cm]{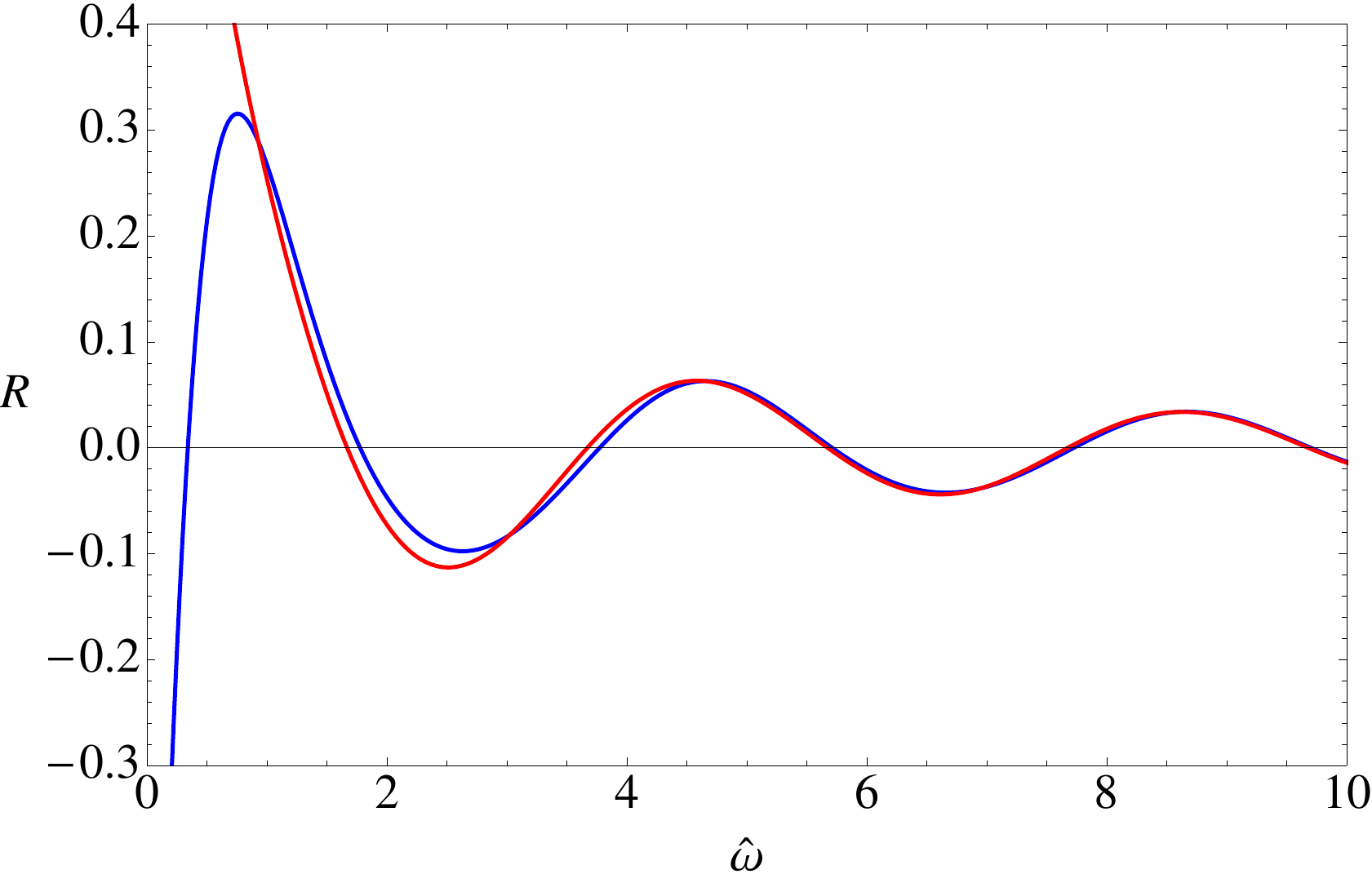}
\caption {The behavior of the full function $R(\hat{\omega},u_s)$ (blue) together with its large-${\hat{\omega}}$ limit $R_\text{asym}({\hat{\omega}},u_s)$ (red) for $r_s/r_h=1.1$.}
\label{Rapprox}
\end{figure}

Inserting finally the straightforwardly obtainable large-$\hat{\omega}$ limit of $c_-/c_+$ to eq.~(\ref{PiTas}), we obtain an analytic, though somewhat cumbersome-looking expression for the retarded correlator,
\ba
&&\Pi^\perp_\text{asym}(\hat{\omega},u_s)=-\frac{N_c^2T^2}{8}\frac{\Gamma(2/3)}{\Gamma(1/3)}\times e^{i 2\pi/3}3^{1/3}\hat{\omega}^{2/3}\label{Piasym} \\
&&\times
\frac{8u_s^{3/2}\hat{\omega}-\sqrt{1-u_s^2}(1+\sqrt{1-u_s^2})(i+e^{i\pi/3}e^{i\hat{\omega} g(u_s)})}{8u_s^{3/2}\hat{\omega}-\sqrt{1-u_s^2}(1+\sqrt{1-u_s^2})(i-e^{i2\pi/3}e^{i\hat{\omega} g(u_s)})} \, . \nonumber
\eea
From here, we easily obtain the large-$\w$ limit of corresponding spectral density, 
\ba
\label{chiasym}
\hspace{-0.7cm}\chi_\text{asym}(\hat{\omega},u_s)&=& \frac{N_c^2T^2}{4}\frac{3^{5/6}\Gamma(2/3)}{\Gamma(1/3)}\nn\\
&\times&\bigg(1+\frac{\sqrt{1-u_s^2}(1+\sqrt{1-u_s^2})}{8u_s^{3/2}}\nn\\
&\times&\frac{\cos(\w g(u_s))+\sqrt{3}\sin(\w g(u_s))}{\w}\nn\\
&+&{\mathcal O}(1/\w^2)\bigg) \, ,
\ea
which is seen to reduce to the correct equilibrium limit as $u_s\to 1$. In fig.~\ref{Rapprox}, where we plot the asymptotic form of $R$, derived using eq.~(\ref{Piasym}), together with our full numerical result, we observe excellent agreement of the two curves already at moderately small frequencies.

{\em Conclusions.}
In the paper at hand, we have investigated the production of prompt photons in strongly coupled, out-of-equilibrium ${\mathcal N}=4$ SYM plasma. This work can be viewed as a generalization of the equilibrium computation of Caron-Huot et~al.~\cite{Huot} to a specific model of holographic thermalization, involving the gravitational collapse of a thin spherical shell in AdS$_5$ spacetime \cite{Danielsson1}. 

Our main results are depicted in figs.~\ref{rho}--\ref{photonrate}. They display a distinctive pattern of fluctuations in the spectral density, which however are significantly dampened in the photo-emission spectrum. In these figures, the thermalization process is parameterized by the radial location of the shell, but as discussed in section 4 of \cite{Rudolf}, explicit time dependence can be introduced at any point by solving the equation of motion of the shell (upon specifying its matter content and initial condition).

The fashion, in which the photon spectral density is seen to approach its thermal limit, is very similar to that computed for dileptons at rest in ref.~\cite{Rudolf}. It is indicative of a top/down type thermalization pattern, and shows no apparent dependence on the virtuality of the produced photons (see cf.~e.g.~\cite{Arnold2} for a different conclusion in the case of jet quenching).


{\em Acknowledgments.}
We thank Nicolas Borghini, Ville Ker\"anen, Esko Keski-Vakkuri, Anton Rebhan, Dominik Steineder and Bin Wu for useful discussions. S.S.,~O.T.~and A.V.~were supported by the Sofja Kovalevskaja program of the Alexander von Humboldt Foundation, and S.S.~additionally by the START project Y435-N16 of the Austrian Science Fund (FWF).



\begin{thebibliography}{99}


\bibitem{Arleo}
  F.~Arleo,
  J.\ Phys.\ G G {\bf 38} (2011) 124017
  [arXiv:1109.3121 [hep-ph]].

\bibitem{Gale}
  C.~Gale,
  arXiv:0904.2184 [hep-ph].
  
  

\bibitem{Adare:2008ab}
  A.~Adare {\it et al.}  [PHENIX Collaboration],
  Phys.\ Rev.\ Lett.\  {\bf 104}, 132301 (2010)
  [arXiv:0804.4168 [nucl-ex]].


\bibitem{Sakaguchi:2010hx}
  T.~Sakaguchi,
  Nucl.\ Phys.\ A {\bf 855}, 141 (2011)
  [arXiv:1012.1893 [nucl-ex]].



\bibitem{Reygers:2009tm}
  K.~Reygers [PHENIX Collaboration],
  arXiv:0908.2382 [nucl-ex].


\bibitem{Tannenbaum}
  M.~J.~Tannenbaum,
  arXiv:1201.5900 [nucl-ex].

\bibitem{Muller}
  B.~Muller, J.~Schukraft and B.~Wyslouch,
  arXiv:1202.3233 [hep-ex].

\bibitem{Arnold}
  P.~B.~Arnold, G.~D.~Moore and L.~G.~Yaffe,
  JHEP {\bf 0112} (2001) 009
  [hep-ph/0111107].

\bibitem{Meyer}
  H.~B.~Meyer,
  Eur.\ Phys.\ J.\ A {\bf 47} (2011) 86
  [arXiv:1104.3708 [hep-lat]].


\bibitem{Maldacena}
  J.~M.~Maldacena,
  Adv.\ Theor.\ Math.\ Phys.\  {\bf 2}, 231 (1998)
  [Int.\ J.\ Theor.\ Phys.\  {\bf 38}, 1113 (1999)]
  [hep-th/9711200].


\bibitem{Gubser}
  S.~S.~Gubser, I.~R.~Klebanov and A.~M.~Polyakov,
  Phys.\ Lett.\ B {\bf 428} (1998) 105
  [hep-th/9802109].

\bibitem{Witten}
  E.~Witten,
  Adv.\ Theor.\ Math.\ Phys.\  {\bf 2}, 253 (1998)
  [hep-th/9802150].



\bibitem{Danielsson1}
  U.~H.~Danielsson, E.~Keski-Vakkuri and M.~Kruczenski,
  Nucl.\ Phys.\ B {\bf 563}, 279 (1999)
  [hep-th/9905227].

\bibitem{Danielsson2}
  U.~H.~Danielsson, E.~Keski-Vakkuri and M.~Kruczenski,
  JHEP {\bf 0002}, 039 (2000)
  [hep-th/9912209].

\bibitem{LinShuryakIII}
  S.~Lin and E.~Shuryak,
  Phys.\ Rev.\ D {\bf 78}, 125018 (2008)
  [arXiv:0808.0910 [hep-th]].
  %


\bibitem{Balasubramanian}
  V.~Balasubramanian, A.~Bernamonti, J.~de Boer, N.~Copland, B.~Craps, E.~Keski-Vakkuri, B.~Muller and A.~Schafer {\it et al.},
  Phys.\ Rev.\ D {\bf 84} (2011) 026010
  [arXiv:1103.2683 [hep-th]].

\bibitem{Rudolf}
  R.~Baier, S.~A.~Stricker, O.~Taanila and A.~Vuorinen,
  JHEP {\bf 1207} (2012) 094
  [arXiv:1205.2998 [hep-ph]].
  
\bibitem{Wu:2012ri} 
  B.~Wu,
  arXiv:1208.1393 [hep-th].

  \bibitem{Huot}
  S.~Caron-Huot, P.~Kovtun, G.~D.~Moore, A.~Starinets and L.~G.~Yaffe,
  JHEP {\bf 0612} (2006) 015
  [hep-th/0607237].

  \bibitem{Myers}
  R.~C.~Myers, A.~O.~Starinets and R.~M.~Thomson,
  JHEP {\bf 0711} (2007) 091
  [arXiv:0706.0162 [hep-th]].


  \bibitem{SonStarinets}
  D.~T.~Son and A.~O.~Starinets,
  JHEP {\bf 0209}, 042 (2002)
  [hep-th/0205051].


\bibitem{Chesler}
  P.~M.~Chesler and D.~Teaney,
  arXiv:1112.6196 [hep-th].

\bibitem{Galante}
  D.~Galante and M.~Schvellinger,
  arXiv:1205.1548 [hep-th].
  
\bibitem{Erdmenger}
  J.~Erdmenger and S.~Lin,
  arXiv:1205.6873 [hep-th].

\bibitem{Caceres}
  E.~Caceres and A.~Kundu,
  arXiv:1205.2354 [hep-th].

\bibitem{Mukhopadhyay}
  A.~Mukhopadhyay,
  arXiv:1206.3311 [hep-th].



\bibitem{Kovtun}
  P.~K.~Kovtun and A.~O.~Starinets,
  Phys.\ Rev.\ D {\bf 72} (2005) 086009
  [hep-th/0506184].



\bibitem{Abramowitz}
  M.~Abramowitz and I.~A.~Stegun,
  Dover Publications, New York, 1964.

\bibitem{Hassanain}
  B.~Hassanain and M.~Schvellinger,
  Phys.\ Rev.\ D {\bf 85} (2012) 086007
  [arXiv:1110.0526 [hep-th]].
   
\bibitem{Arnold2}
  P.~Arnold and D.~Vaman,
  JHEP {\bf 1104} (2011) 027
  [arXiv:1101.2689 [hep-th]].

\end{thebibliography}
\end{document}